\documentclass{article}
\usepackage{spconf,amsmath,graphicx,multirow,float,hyperref,amssymb}
\usepackage{booktabs}
\usepackage{arydshln}
\usepackage{xcolor}
\usepackage{adjustbox}
\usepackage{enumitem}

\usepackage[labelformat=simple]{subcaption}

\usepackage[font=small,labelfont=bf]{caption}

\usepackage[subtle,title=normal]{savetrees}

\usepackage[T1]{fontenc}

\newlength{\bibitemsep}
\newlength{\bibparskip}
\let\oldthebibliography\thebibliography
\renewcommand\thebibliography[1]{%
  \oldthebibliography{#1}%
  \setlength{\parskip}{\bibitemsep}%
  \setlength{\itemsep}{\bibparskip}%
}

\title{Adapting self-supervised models to multi-talker speech recognition \\using speaker embeddings}
\name{Zili Huang, Desh Raj, Paola Garc\'ia, Sanjeev Khudanpur}
\address{Center for Language and Speech Processing and Human Language Technology Center of Excellence,\\Johns Hopkins University, Baltimore, USA}
\begin{document}
\ninept
\maketitle
\begin{abstract}
Self-supervised learning (SSL) methods which learn representations of data without explicit supervision have gained popularity in speech-processing tasks, particularly for single-talker applications. However, these models often have degraded performance for multi-talker scenarios --- possibly due to the domain mismatch --- which severely limits their use for such applications. In this paper, we investigate the adaptation of upstream SSL models to the multi-talker automatic speech recognition (ASR) task under two conditions. First, when segmented utterances are given, we show that adding a \textit{target speaker extraction} (TSE) module based on enrollment embeddings is complementary to mixture-aware pre-training. Second, for unsegmented mixtures, we propose a novel \textit{joint speaker modeling} (JSM) approach, which aggregates information from all speakers in the mixture through their embeddings. With controlled experiments on Libri2Mix, we show that using speaker embeddings provides relative WER improvements of 9.1\% and 42.1\% over strong baselines for the segmented and unsegmented cases, respectively. We also demonstrate the effectiveness of our models for real conversational mixtures through experiments on the AMI dataset.
\end{abstract}
\vspace{-1em}
\section{Introduction}
\label{sec:intro}
\vspace{-.5em}
Self-supervised learning (SSL), which obtains supervisory signals from the input data itself, is an active sub-area of unsupervised learning~\cite{mohamed2022self,liu2021self}. Generally speaking, the SSL pipeline consists of a \textit{pre-training stage}, in which an upstream model is trained on a large quantity of unlabeled data with some pretext tasks, and a \textit{fine-tuning stage}, in which the upstream model is adapted for specific downstream tasks with annotated data.

SSL has recently gained popularity in the speech community due to its strong performance and low adaptation cost. By simply replacing the acoustic features with hidden representations from pre-trained upstream models~\cite{yang2021superb,chen2022wavlm}, or fine-tuning the entire model with additional task-specific layers~\cite{baevski2020wav2vec,hsu2021hubert}, SSL models have surpassed state-of-the-art downstream-only baselines for several tasks including speech recognition~\cite{baevski2020wav2vec,hsu2021hubert} and speaker recognition~\cite{chen2022large,fan2020exploring}. Furthermore, these pre-trained upstream models often require very small amounts of adaptation data; for instance, wav2vec 2.0 achieves a word error rate (WER) of 8.2\% on LibriSpeech~\cite{Panayotov2015LibrispeechAA} \texttt{test-other} after fine-tuning with 10 minutes of labeled speech~\cite{baevski2020wav2vec}.

\begin{figure}[t]
    \centering
    \includegraphics[width=\linewidth]{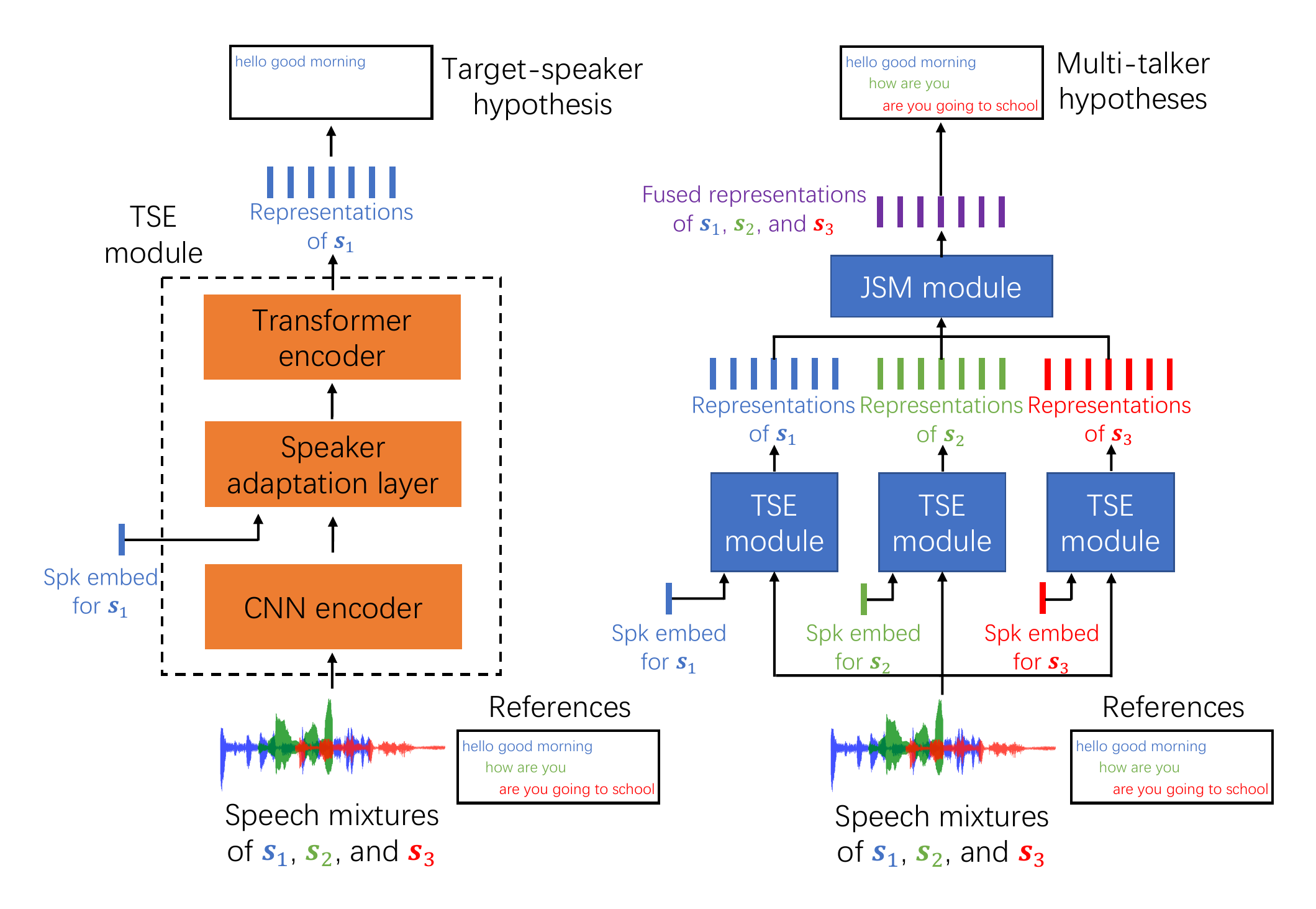}
    \vspace{-2em}
    \caption{Our proposed Target Speaker Extraction (TSE) module (\textit{left}) and Joint Speaker Modeling (JSM) module (\textit{right}) for adapting SSL-based upstream models to multi-talker ASR.}
    \label{fig:model}
    \vspace{-1.5em}
\end{figure}


Despite their strong performance for various downstream tasks, most of the existing SSL models only focus on single-talker scenarios. The previous studies on pre-training and fine-tuning upstream SSL models for multi-talker scenarios are very limited~\cite{chen2022wavlm,chang2022end}. Such scenarios are usually more challenging due to the involvement of multiple speakers, quick turn-taking, and overlapping speech~\cite{ccetin2006analysis}.


In this paper, we focus on the fine-tuning stage of the SSL pipeline and investigate the adaptation of SSL-based upstream models to the multi-talker ASR task, for both segmented utterances and unsegmented mixtures. For the former, we build upon previous work on target speaker extraction and recognition~\cite{delcroix2018single,vzmolikova2019speakerbeam,medennikov2020target,wang2018voicefilter}; we use target speaker embeddings (extracted from enrollment audio) to direct the model to transcribe a specific speaker in the segmented utterance containing mixed speech. We perform a thorough analysis for several modeling aspects, including embedding type and dimensionality, fusion methods for the adaptation layer, and the effect of upstream models. For the case of unsegmented mixtures, we extend the TSE approach by iteratively decoding each speaker in the mixtures, and we also propose a novel \textit{joint speaker modeling} (JSM) technique that aggregates information from all speakers through their embeddings, and transcribes them simultaneously.
 
Our contributions are summarized as follows: (1) we propose methods to adapt SSL models to the multi-talker ASR task based on the speaker embeddings; (2) we conduct a detailed analysis of different modeling aspects of target speaker extraction with SSL models; and (3) we evaluate our methods on both artificial and real datasets, namely LibriMix and AMI. Our code will be released at the time of publication.

\vspace{-1em}
\section{The multi-talker ASR problem}
\vspace{-.5em}
\label{sec:problem}
In conventional ASR, the input is a single-talker, segmented utterance $\mathbf{x}$ (in time or time-frequency domain), and the system is required to predict hypothesis $\mathbf{y} = f(\mathbf{x};\Theta_f)$, where $\mathbf{y}$ is a token sequence and $f$ is a function parametrized by $\Theta_f$ such as a neural network. In multi-talker ASR, the input is an unsegmented mixture $\mathbf{m}$ containing speech from multiple speakers (say, $K$ speakers), i.e., $\mathbf{m} = \sum_{u=1}^U \mathbf{x}_u$, where $\mathbf{x}_u$ is the $u$-th utterance and is shifted and zero-padded to the length of $\mathbf{m}$. Here, we assume that the additive noise and convolutive reverberations are included in the individual terms. Additionally, we may be provided enrollment embeddings $\mathbf{e}_k \in \mathbb{R}^d$ for the $K$ speakers in the mixture. The desired output is $\mathbf{y} = \bigcup_{u=1}^U \mathbf{y}_u$, where $\mathbf{y}_u$ is the transcription corresponding to $\mathbf{x}_u$. Depending on whether the mixture is segmented, two cases are possible.

\vspace{-1em}
\subsection{Segmented utterances}
\label{sec:segmented}
\vspace{-0.5em}

In this easier case, we assume that the spans and speaker identities for all $U$ utterances are known, such that $\mathbf{y}_u$ can be directly estimated: $\widehat{\mathbf{y}}_u = g_{\mathrm{S}}(\mathbf{m}, \Delta_u, \mathbf{e}_{k_u};\Theta_{g_{\mathrm{S}}})$, where $\Delta_u$ denotes the span of $\mathbf{x}_u$ in the mixture, $k_u$ is the speaker index corresponding to utterance $\mathbf{x}_u$. By repeating this inference $U$ times, we get the desired multi-talker hypothesis $\widehat{\mathbf{y}}$. In practice, $\Delta_u$ and $k_u$ may be obtained using speaker diarization.


\vspace{-1em}
\subsection{Unsegmented mixture}
\label{sec:unsegmented}
\vspace{-0.5em}

In this case, the individual segmentation of $\mathbf{x}_u$ is not known in advance, therefore utterance-level transcription is not feasible. If we assume that speakers are enrolled, we can instead perform \textit{speaker-level} transcription, i.e., $\widehat{\mathbf{y}}^{\prime} = \bigcup_{k=1}^K \widehat{\mathbf{y}}^k$, where $\widehat{\mathbf{y}}^k$ denotes the hypothesis for speaker $k$. This may be done by \textit{iteratively} decoding each speaker using a same single-talker parametrized model, i.e.,
\begin{equation}
    \widehat{\mathbf{y}}^k = g_\mathrm{S}(\mathbf{m}, \Delta, \mathbf{e}_k; \Theta_{g_{\mathrm{S}}}),
\end{equation}
where $\Delta$ is a dummy span indicating the entire duration of $\mathbf{m}$. Alternatively, we can define a multi-talker parametrized model $g_\mathrm{M}$ and perform \textit{joint} decoding of all speakers, i.e.,
\begin{equation}
    \widehat{\mathbf{y}}^{\prime} = g_\mathrm{M}(\mathbf{m}, \mathbf{E}_{1:K}; \Theta_{g_{\mathrm{M}}}),
\end{equation}
where $\mathbf{E}\in \mathbb{R}^{d\times K}$ contains all $K$ speaker embeddings.

In the next section, we will describe our proposed methods to use self-supervised models for both $g_{\mathrm{S}}$ and $g_{\mathrm{M}}$, so that we can handle segmented and unsegmented inputs.

\vspace{-1em}
\section{Adapting self-supervised models with speaker embeddings}
\label{sec:methodology}
\vspace{-.5em}
First, we will show how to adapt self-supervised models for the case of single-talker transcription (i.e., $g_{\mathrm{S}}$) using speaker embeddings. This model can be used for segmented utterances or iterative decoding of unsegmented mixtures. Thereafter, we will develop a multi-talker extension (i.e., $g_{\mathrm{M}}$) of the model to perform joint decoding of all speakers in the mixture. Since our methods rely on the architecture of the upstream SSL models, we will briefly describe them here.

We used three variants of the wav2vec 2.0~\cite{baevski2020wav2vec} class of SSL models: HuBERT~\cite{hsu2021hubert}, WavLM Base, and WavLM Base+~\cite{chen2022wavlm}. These models consist of convolutional and Transformer encoders. For all models, the convolutional encoder contains 7 layers that down-sample the audio input by a factor of 320, such that the extracted features (say $\mathbf{h}_{\mathrm{conv}}$) have a stride of 20ms and a receptive field of 25ms. The CNN encoder is followed by a 12-block Transformer encoder, which produces contextual output representations $\mathbf{h}_{\mathrm{trf}}$. Each Transformer block consists of a 768-dim self-attention layer with 8 heads and a 3,072-dim feed-forward layer. WavLM additionally uses relative gated positional bias in the self-attention layers~\cite{Chi2022XLMECL}. 

\vspace{-.8em}
\subsection{Target speaker extraction (TSE)}
\label{ssec:TSE}
\vspace{-.5em}

We adapt SSL models for single-talker transcription through a ``target speaker extraction'' (TSE) component that models $g_{\mathrm{S}}$, as illustrated in Fig.~\ref{fig:model} (\textit{left}). 
Specifically, we insert a \textit{speaker adaptation layer} between the CNN and Transformer blocks of the encoder. This layer adapts $\mathbf{h}_{\mathrm{conv}}$ using the speaker embedding $\mathbf{e}_k$ before feeding them to the Transformer encoder. By fine-tuning the model with multi-talker mixtures, speaker embeddings, and corresponding references, the adaptation layer learns to extract only target-speaker features from the CNN representations. The adapted CNN representations, $\widehat{\mathbf{h}}_{\mathrm{conv}}$, are passed to the Transformer encoder which produces speaker-specific contextual representations, $\widehat{\mathbf{h}}_{\mathrm{trf}}$. A linear projection layer maps $\widehat{\mathbf{h}}_{\mathrm{trf}}$ to the output dimensionality, and the CTC loss~\cite{graves2006connectionist} is used for end-to-end training.

In our experiments, we tried four different methods for the implementation of the speaker adaptation layer.

\begin{enumerate}[leftmargin=*,topsep=0pt]
    \item \textbf{ADD}: $\mathbf{e}_{k}$ is added to $\mathbf{h}_{\mathrm{conv}}$ through a linear projection. 

    \item \textbf{CAT}: $\mathbf{e}_k$ is appended to $\mathbf{h}_{\mathrm{conv}}$ and the concatenated representation is projected back to the original dimension through a linear projection.

    \item \textbf{FiLM}: Feature-wise linear modulation~\cite{perez2018film} estimates a feature-wise affine transformation based on the conditioning information, i.e., $\widehat{\mathbf{h}}_{\mathrm{conv}} = \gamma(\mathbf{e}_k) \cdot \mathbf{h}_{\mathrm{conv}} + \beta(\mathbf{e}_k)$, where $\gamma$ and $\beta$ can be arbitrary functions. In our experiment, $\gamma$ and $\beta$ are two linear layers.


    
    \item \textbf{CLN}: In standard Transformer layers, layer normalization~\cite{ba2016layer} has been applied to normalize the distributions of intermediate layers, formulated as
    \begin{equation}
        \widehat{\mathbf{h}} = \frac{\mathbf{h}-\mathbf{\mu}}{\mathbf{\sigma}} \cdot \gamma + \beta
    \end{equation}
    where $\mathbf{\mu}$ and $\mathbf{\sigma}$ are the mean and standard deviation of input features $\mathbf{h}$, $\gamma$ and $\beta$ are learnable scaling and shifting parameters for the affine transformation. Conditional layer normalization (CLN)~\cite{pilault2020conditionally} replaces $\gamma$ with a speaker-specific scaling $\widehat{\gamma}(\mathbf{e}_{k})$, i.e., 
    \begin{equation}
         \widehat{\gamma}(\mathbf{e}_{k}) = w(\mathbf{e}_{k}) \cdot \gamma + b(\mathbf{e}_{k})
    \end{equation}
    Basically, it transforms $\gamma$ with a FiLM layer conditioned on the speaker embeddings $\mathbf{e}_{k}$. Different from the three above methods which take $\mathbf{h}_{\mathrm{conv}}$ as input, CLN is introduced to the bottom Transformer block by replacing the original two LNs.


\end{enumerate}




\vspace{-1em}
\subsection{Joint speaker modeling (JSM)}
\label{ssec:JSM}
\vspace{-.5em}

In principle, the TSE module can be used for the transcription of unsegmented mixtures by iterating over all speakers. However, this setup may be sub-optimal for multi-talker ASR since the hypothesis produced for a speaker is independent of other speakers. Some errors, like repetitive decoding of the same utterance, could be potentially eliminated if the model has an overview of all speakers presented in the speech. Furthermore, the TSE approach would require $K$ decoding iterations, resulting in higher inference time. Inspired by the architecture of target-speaker voice activity detection (TS-VAD)~\cite{medennikov2020target}, we propose ``joint speaker modeling'' (JSM) as illustrated in Fig.~\ref{fig:model} (\textit{right}). We first use the TSE module to extract speaker-specific contextual representations $\mathbf{h}_{\mathrm{trf}}^k$ for all speakers in the speech mixture $\mathbf{m}$. Thereafter, the JSM module models $g_{\mathrm{M}}$ by computing $\widehat{\mathbf{y}}^{\prime}$ directly using $\{\mathbf{h}_{\mathrm{trf}}^k|k=1,2,...,K\}$. In our implementation, the representations $\mathbf{h}_{\mathrm{trf}}^k$ are concatenated along the feature axis and projected back to the original dimension. The combined input then pass through a Transformer layer to get multi-talker contextual representations $\widehat{\mathbf{h}}_{\mathrm{trf}}$. Finally, a linear layer maps $\widehat{\mathbf{h}}_{\mathrm{trf}}$ to $K\times |V|$ output units (where $|V|$ is vocabulary size) to get token-level posteriors for each speaker. The model is trained with the sum of CTC losses over all $K$ speakers.


\vspace{-1em}
\section{Experimental Setup}
\label{sec:exp}
\vspace{-.5em}

\subsection{Data}
\vspace{-.5em}
We evaluated our method on LibriMix~\cite{cosentino2020librimix} and AMI~\cite{carletta2005ami} dataset. LibriMix is a synthetic dataset consisting of simulated speech mixtures.
In our experiments, we used the ``2-spk 16kHz max'' condition (hereby denoted as Libri2Mix). The training set consisted of 58 hours of speech mixtures created from LibriSpeech~\cite{Panayotov2015LibrispeechAA} \texttt{train-clean-100} while the \texttt{dev} and \texttt{test} set contained 11 hours each.

AMI~\cite{carletta2005ami} is a meeting-style corpus consisting of $\sim$100h of annotated audio/video recordings. Each meeting has 3 to 5 participants, and was recorded using headset microphones and two 8-ch microphone arrays. In our experiments, we use the so-called ``IHM-mix" evaluation condition, which is generated by mixing all headset recordings from the same meeting. The resulting audios contain a significant amount of overlap but no far-field artifacts, allowing us to evaluate our models specifically for overlapping speech.

Our proposed methods require enrollment utterances for speaker embedding extraction. For Libri2Mix, we collected 15 seconds of enrollment speech per speaker from the Libri-Light~\cite{librilight} project (excluding the LibriSpeech portion). For AMI IHM-mix, since the same speakers participate in multiple meetings, we used clean segments from other meetings as enrollment utterances. For example, if the speaker \texttt{FEE041} participated in four meetings \texttt{ES2011{a-d}}, the clean segments in meetings \texttt{ES2011{b-d}} would be considered for enrollment of \texttt{FEE041} in meeting \texttt{ES2011a}.

\vspace{-1em}
\subsection{Evaluation metrics}
\vspace{-0.5em}

\begin{figure}[t]
    \centering
    \includegraphics[width=1.0\linewidth]{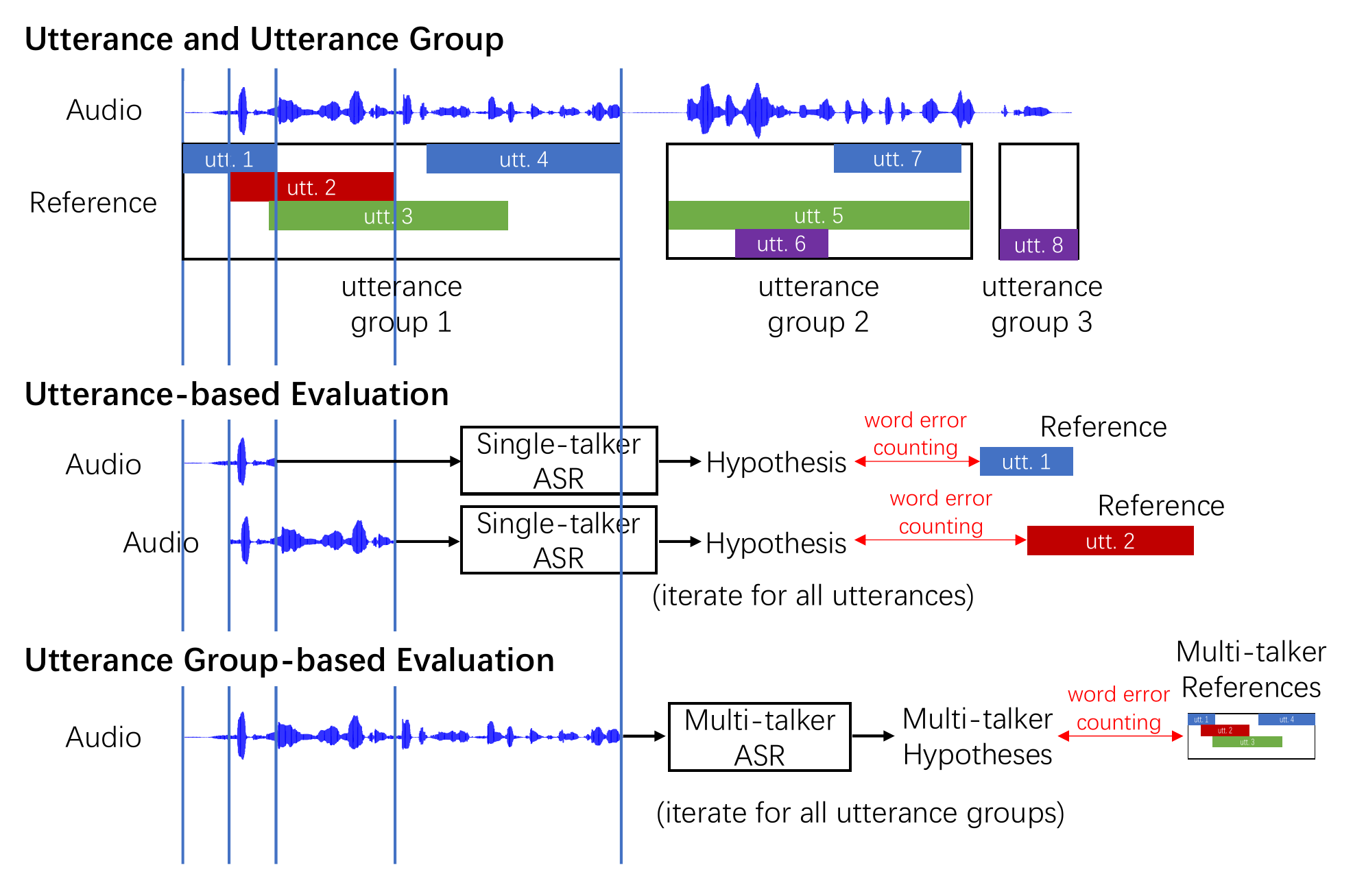}
    \caption{Utterance-based and utterance group-based evaluation schemes for segmented utterances and unsegmented mixtures, respectively.}
    \label{fig:evaluation}
    \vspace{-1.5em}
\end{figure}

The two tasks with segmented utterances (\S~\ref{sec:segmented}) and unsegmented mixtures (\S~\ref{sec:unsegmented}) require different evaluation schemes. Following \cite{kanda2021large}, we refer to these as \textit{utterance-based} and \textit{utterance group-based} evaluation, respectively, which are illustrated in Fig.~\ref{fig:evaluation}. 

In utterance-based evaluation, the exact time boundary for each speaker segment is known. We perform inference on each speaker segment, and our hypothesis is compared with the reference to compute the word error rate (WER). 
To evaluate unsegmented mixtures, we defined an ``utterance group'' as a set of utterances connected by speaker overlaps. In practice, a voice activity detector (VAD) may be used to estimate utterance group boundaries. 
If the correspondence between the multi-talker hypotheses and references in an utterance group is unknown (as is the case in \cite{kanda2021large}), concatenated minimum-permutation WER (cpWER)~\cite{watanabe2020chime} may be computed instead of WER.
In our proposed model, the output order of hypotheses is determined by the order of speaker embeddings, therefore we can directly compute the WER between the hypotheses and references.


\vspace{-1em}
\subsection{Model configuration and baselines}
\vspace{-.5em}
The upstream models we used are all pre-trained with the same masked cluster index prediction loss~\cite{hsu2021hubert}. However, while HuBERT is trained only on single-talker audios, WavLM additionally uses simulated mixtures for pre-training, which makes it more suitable for multi-talker downstream tasks. HuBERT and WavLM Base are pre-trained using LibriSpeech (960h), while WavLM Base+ uses 94k hours of mixtures comprising large-scale diverse data, such as LibriLight~\cite{kahn2020libri},
VoxPopuli~\cite{wang2021voxpopuli}, and GigaSpeech~\cite{GigaSpeech2021}.

Besides the upstream model, the choice of pre-trained speaker embedding extractor and the dimensionality of $\mathbf{e}_k$ is also important towards the final performance. We compared the multi-talker ASR performance for three different embedding methods --- i-vectors~\cite{dehak2010front}, x-vectors~\cite{snyder2018x}, and ECAPA-TDNN embeddings~\cite{desplanques2020ecapa}. I-vectors are unsupervised embeddings that model the total variability between an utterance and the universal background model. X-vectors~\cite{snyder2018x} and ECAPA-TDNN embeddings, on the other hand, are extracted from discriminatively-trained neural networks after pooling layers. While the former uses conventional time-delay neural networks (TDNNs)~\cite{Peddinti2015ATD}, the latter applies several enhancements on the frame and segment pooling components through squeeze-and-excitation~\cite{Hu2020SqueezeandExcitationN} and channel-dependent attention, respectively. For speaker recognition, DNN-based embeddings have been found to outperform i-vectors; ECAPA-TDNN, in particular, has obtained state-of-the-art performance on the VoxCeleb benchmark~\cite{desplanques2020ecapa}. For our experiments, all these three embedding extractors were pre-trained on the VoxCeleb~\cite{nagrani2017voxceleb,chung2018voxceleb2} dataset, with a dimension of 400, 512, and 256 for i-vectors, x-vectors, and ECAPA-TDNN embeddings, respectively.

For segmented utterances, we compared our models with a baseline that does not use any speaker embeddings but is otherwise fine-tuned on the same data. For unsegmented mixtures, we implemented a permutation invariant training (PIT)~\cite{yu2017permutation} baseline as follows. We added a linear layer consisting of $K\times |V|$ output units on top of the pre-trained SSL model, where each of the $K$ sets is expected to transcribe a distinct speaker. The model was then trained using PIT, i.e., using the minimum sum of losses between all permutations of reference and hypotheses speakers. Since this model does not have a deterministic speaker ordering, we evaluated it using cpWER instead of regular WER.

\vspace{-1em}
\section{Results \& Discussion}
\label{ssec:main}
\vspace{-.5em}

\begin{table}[t]
    \centering
    \caption{WER (\%) on Libri2Mix and AMI IHM-mix for utterance-based evaluations. TSE used x-vectors fused by the CLN approach.}
    \adjustbox{max width=\linewidth}{
    \begin{tabular}{@{}lcc@{}}
    \toprule
        \textbf{Model} & \textbf{Libri2Mix} & \textbf{AMI} \\
        \midrule
        WavLM Base+ & 8.34 & 18.97 \\ 
        ~~ + TSE & \textbf{7.58} & \textbf{18.07} \\
    \bottomrule
    \end{tabular}}
    \label{tab:main-seg}
\end{table}

\begin{figure}[t]
    \centering
    \includegraphics[width=0.8\linewidth]{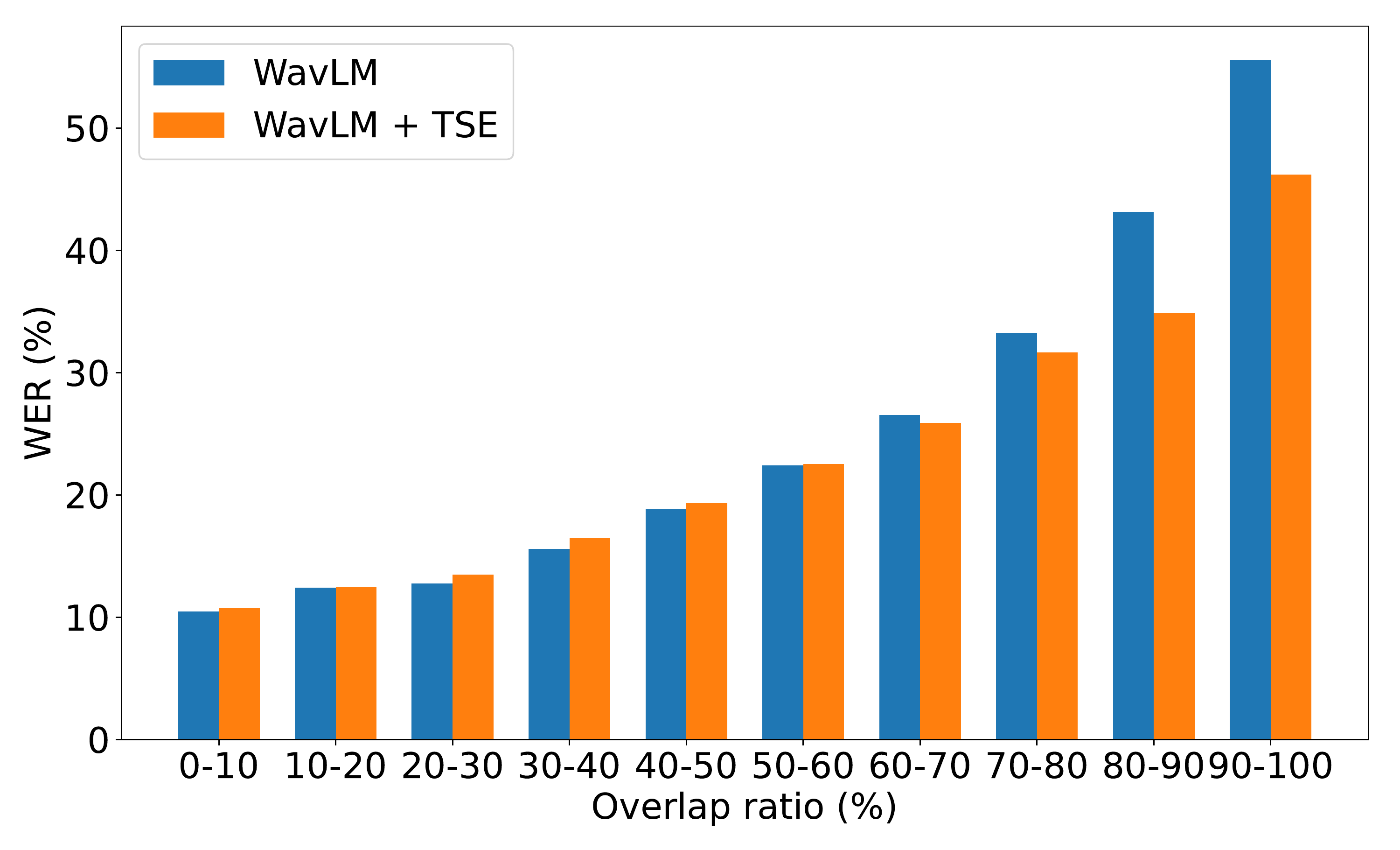}
    \vspace{-1em}
    \caption{WER decomposition by overlap ratio for AMI IHM-mix.}
    \label{fig:overlapratio}
    \vspace{-2em}
\end{figure}

\begin{table*}[t]
    \begin{minipage}{.29\linewidth}
      \caption{WER (\%) on Libri2Mix for different SSL-based upstream models with and without the TSE module.}
    \label{tab:upstream}
      \vspace{-0.8em}
      \centering
      \adjustbox{max width=\linewidth}{%
    \begin{tabular}{@{}lcc@{}}
    \toprule
        \textbf{Model} & \textbf{w/o TSE} & \textbf{w/ TSE} \\
        \midrule
        HuBERT Base & 13.04 & 11.62 {\scriptsize(\textdownarrow 10.9\%)} \\
        WavLM Base & 9.81 & 9.03 {\scriptsize(\textdownarrow 8.0\%)} \\
        WavLM Base+ & \textbf{8.34} & \textbf{7.58} {\scriptsize(\textdownarrow 9.1\%)} \\
        \bottomrule
    \end{tabular}}
    \end{minipage}%
    \hfill
    \begin{minipage}{.37\linewidth}
      \centering
        \caption{WER (\%) on Libri2Mix for different types and dimensionality of speaker embeddings.}
        \label{tab:spk_embed}
        \vspace{-0.8em}
        \adjustbox{max width=\linewidth}{%
    \begin{tabular}{@{}lcc@{\hskip 20pt}cc@{}}
    \toprule
        \multirow{2}{*}{\textbf{Embed. type}} & \multicolumn{2}{c}{\textbf{CLN}} & \multicolumn{2}{c}{\textbf{ADD}} \\
        \cmidrule(r{20pt}){2-3} \cmidrule(l{0pt}){4-5}
        & $d=64$ & $d=128$ & $d=64$ & $d=128$ \\
        \midrule
        i-vector & 7.82 & 8.04 & 8.25 & 8.54 \\
        ECAPA-TDNN & 7.87 & 8.14 & 8.00 & 8.65 \\ 
        x-vector & \textbf{7.58} & \textbf{7.65} & \textbf{7.86} & \textbf{7.91} \\
        \bottomrule
    \end{tabular}}
        \end{minipage}%
        \hfill
    \begin{minipage}{.28\linewidth}
      \centering
        \caption{WER (\%) on Libri2Mix for different fusion methods in the speaker adaptation layer.}
        \label{tab:fusion_methods}
        \vspace{-1em}
        \adjustbox{max width=\linewidth}{%
    \begin{tabular}{@{}ccccc@{}}
    \toprule
        \multirow{2}{*}{\textbf{Method}} & \multicolumn{2}{c}{\textbf{utterance}} & \multicolumn{2}{c}{\textbf{utterance group}} \\
     \cmidrule(r{2pt}){2-3} \cmidrule(l{2pt}){4-5}
        & \texttt{dev} & \texttt{test} & \texttt{dev} & \texttt{test} \\
      \midrule
        ADD & 7.52	& 7.63 & 12.13 & 14.28 \\
        CAT & 7.45 & 7.65 & 11.96 &	13.33 \\
        FiLM & 7.31 & 7.66 & 12.45 & 13.84 \\
        CLN & \textbf{7.09} & \textbf{7.58} & \textbf{11.04} & \textbf{12.32} \\ 
    \bottomrule
    \end{tabular}}
    \end{minipage}
    \vspace{-1.5em}
\end{table*}

\vspace{-.5em}
\subsection{Utterance-based evaluation of segmented utterances}
\vspace{-.5em}
We evaluated the impact of our proposed TSE module on Libri2Mix and AMI IHM-Mix using WavLM Base+, with x-vector embeddings fused by the CLN approach. These results are shown in Table \ref{tab:main-seg}. For utterance-based evaluation, the baseline WavLM system worked well even without any enrollment information. Despite the high overlap ratio of Libri2Mix, it achieved 8.34\% WER on the \texttt{test} set. Similarly, it obtained a competitive WER of 18.97\% on AMI.
Adding the TSE module improved the WERs by 9.1\% and 4.7\% relative on Libri2Mix and AMI, respectively. To better understand where the improvements came from, we analyzed the AMI WER across different overlap ratios, as shown in Fig.~\ref{fig:overlapratio}. Compared with the WavLM baseline, our TSE-enhanced model achieved comparable or slightly worse WER when the overlap ratio was low (below 50\%). However, it \textbf{outperformed WavLM by a clear margin for higher overlap ratios}. 
We conjecture that for lower overlap ratios, WavLM is able to perform well because it transcribes the majority speaker in the utterance. For higher overlap ratios, it cannot predict the target speaker based on the majority duration, and TSE provides this additional information for such cases.


We further conducted several experiments on Libri2Mix to analyze the impact of different factors such as the upstream model, the embedding type and its dimensionality, and the fusion technique used in the speaker adaptation layer. These are described in the following sections.

\vspace{-1em}
\subsubsection{Effect of upstream models}
\vspace{-.5em}


During pre-training, the upstream model learns generalizable task-agnostic representations using SSL, the quality of which usually determines the performance of downstream tasks. In Table~\ref{tab:upstream}, we compared the multi-talker ASR performance for the three upstream models described earlier, using CLN-based fusion with x-vector speaker embeddings.

First, we observe that the WavLM models significantly outperform HuBERT, both with and without the TSE module, due to their enhanced speaker modeling ability and robustness to noise and speaker overlaps. 
Pre-training on a large-scale diverse dataset (WavLM Base+) provided the best WERs, as expected. Nevertheless, adding TSE provides consistent improvements of 8-11\% across all models, indicating that explicit \textbf{speaker enrollment is complementary to mixture-based pre-training} scheme. For the following experiments, we used WavLM Base+ as the upstream model.


\vspace{-1em}
\subsubsection{Effect of speaker embeddings}
\vspace{-.5em}



In this section, we investigate the impact of speaker embedding type and dimensionality. For a fair comparison, we projected the speaker embeddings to the same dimension using principal component analysis (PCA)~\cite{pearson1901liii}. From Table \ref{tab:spk_embed}, we first observe that 64-dim embeddings outperform their 128-dim counterparts, indicating that there may be \textbf{extraneous information encoded in higher dimensions}. Next, even though ECAPA-TDNNs achieve SOTA performance on speaker recognition, they are outperformed by x-vectors when used in the TSE module (for both CLN and ADD fusion).

\vspace{-1em}
\subsubsection{Effect of fusion methods}
\label{ssec:fusion methods}
\vspace{-.5em}


In \S~\ref{ssec:TSE}, we described four different fusion methods for the speaker adaptation layer. Recall that ADD and CAT directly fuse the speaker embedding $\mathbf{e}_k$ to the convolutional representation $\mathbf{h}_{\mathrm{conv}}$, whereas FiLM and CLN estimate affine transforms based on $\mathbf{e}_k$. 
Here, we compare their performance in both utterance-based and utterance group-based evaluation settings, as shown in Table~\ref{tab:fusion_methods}. On utterance-based evaluation, all these four methods have similar performance, with CLN slightly outperforming others. However, when the segmentation is unknown (utterance group-based evaluation), \textbf{CLN outperforms other methods by a large margin}. 






\vspace{-1em}
\subsection{Utterance group-based evaluation of unsegmented mixtures}
\vspace{-.5em}

\begin{table}[t]
    \centering
    \caption{WER (\%) on Libri2Mix and AMI IHM-mix for utterance group-based evaluations of unsegmented mixtures.$^\dagger$PIT does not have a fixed speaker order, therefore the results are shown using cpWER.}
    \adjustbox{max width=\linewidth}{
    \begin{tabular}{@{}lcc@{}}
    \toprule
        \textbf{Model} & \textbf{Libri2Mix} & \textbf{AMI} \\
        \midrule
        WavLM Base+  &104.44 & 100.26 \\
        ~~ + PIT$^\dagger$  & 18.45 & \textbf{27.02} \\
        ~~ + iterative TSE  & 12.32 & 49.47 \\ 
        ~~ + TSE + JSM  & \textbf{10.68} & 28.40 \\
    \bottomrule
    \end{tabular}}
    \label{tab:main-unseg}
    \vspace{-1.8em}
\end{table}


From the above ablation experiments, we demonstrated that the combination of WavLM Base+ upstream model, x-vector speaker embedding, and the CLN fusion method provided the best WER performance.
Next, we used this configuration to conduct utterance group-based evaluations for unsegmented mixtures, and the results are shown in Table~\ref{tab:main-unseg}. 

It is immediately clear that the single-talker WavLM baseline, which performed well on segmented utterances, failed miserably when the segmentation was not known, resulting in WERs above 100\%. This is because, without any additional information, the model does not know which speaker to transcribe in the unsegmented mixtures.
Joint multi-talker inference using permutation invariant training (PIT) helped the model overcome this limitation and the resulting system obtained decent results on the task, achieving a cpWER of 18.45\% on Libri2Mix and 27.02\% on AMI. Recall that since PIT does not have a deterministic speaker order, we use cpWER instead of WER to evaluate its performance.


Next, we evaluated an iterative decoding approach (described in \S~\ref{sec:unsegmented}) based on the TSE module. While this model conveniently outperformed PIT on Libri2Mix with a WER of 12.32\%, its performance on AMI IHM-Mix was sub-par. 
Our error analysis revealed that this was because on AMI IHM-Mix, the model sometimes mistakenly transcribed other speakers when the target speaker didn't talk or only talked for a short period in the utterance group.
We had conjectured that \textbf{this problem of repeated decoding can be alleviated using the JSM module} described in \S~\ref{ssec:JSM}.
As expected, the multi-talker ASR performance was significantly improved after we jointly modeled the activities of all speakers using the JSM module. WERs improved from 12.32\% to 10.68\% on Libri2Mix, and from 49.47\% to 28.40\% on AMI IHM-mix. Although this WER may seem slightly worse than PIT, it is simply because the evaluation of PIT through cpWER is more forgiving. If we discard the speaker ordering in JSM and compute the best permutation word error rate, JSM will outperform PIT with a cpWER of 25.88\%.

\vspace{-1em}
\section{Conclusion}
\vspace{-1em}
\label{sec:conclusion}
In this paper, we investigated the adaptation of self-supervised models to the multi-talker ASR task using speaker embeddings, for both segmented and unsegmented inputs. For the former, we explored a target speaker extraction module that was found to be complementary to mixture-aware pre-training. For the latter, we proposed a novel joint speaker modeling approach to jointly decode all speakers in the mixture. Furthermore, we performed controlled experiments to study the impact of several modeling aspects in target speaker extraction with SSL models. Our experimental results on LibriMix and AMI demonstrated that self-supervised models can achieve substantially better multi-talker ASR performance by utilizing information from speaker embeddings. We will explore the possibility of applying our methods to other multi-talker tasks such as speaker diarization and speech separation in the future.



\small
\bibliographystyle{IEEEbib}
\bibliography{refs}

\begin{thebibliography}{10}

\bibitem{mohamed2022self}
Abdelrahman Mohamed et~al.,
\newblock ``Self-supervised speech representation learning: A review,''
\newblock {\em IEEE Journal of Selected Topics in Signal Processing}, 2022.

\bibitem{liu2021self}
Xiao Liu et~al.,
\newblock ``Self-supervised learning: Generative or contrastive,''
\newblock {\em IEEE Transactions on Knowledge and Data Engineering}, 2021.

\bibitem{yang2021superb}
Shu-wen Yang et~al.,
\newblock ``{SUPERB}: Speech processing universal performance benchmark,''
\newblock in {\em InterSpeech}, 2021.

\bibitem{chen2022wavlm}
Sanyuan Chen et~al.,
\newblock ``{WavLM}: Large-scale self-supervised pre-training for full stack
  speech processing,''
\newblock {\em IEEE Journal of Selected Topics in Signal Processing}, 2022.

\bibitem{baevski2020wav2vec}
Alexei Baevski, Yuhao Zhou, Abdelrahman Mohamed, and Michael Auli,
\newblock ``{Wav2vec 2.0}: A framework for self-supervised learning of speech
  representations,''
\newblock {\em NeurIPS}, 2020.

\bibitem{hsu2021hubert}
Wei-Ning Hsu et~al.,
\newblock ``{HuBERT}: Self-supervised speech representation learning by masked
  prediction of hidden units,''
\newblock {\em IEEE/ACM TASLP}, 2021.

\bibitem{chen2022large}
Zhengyang Chen et~al.,
\newblock ``Large-scale self-supervised speech representation learning for
  automatic speaker verification,''
\newblock in {\em IEEE ICASSP}, 2022.

\bibitem{fan2020exploring}
Zhiyun Fan et~al.,
\newblock ``Exploring wav2vec 2.0 on speaker verification and language
  identification,''
\newblock in {\em InterSpeech}, 2021.

\bibitem{Panayotov2015LibrispeechAA}
Vassil Panayotov et~al.,
\newblock ``Librispeech: An asr corpus based on public domain audio books,''
\newblock in {\em IEEE ICASSP}, 2015.

\bibitem{chang2022end}
Xuankai Chang et~al.,
\newblock ``End-to-end integration of speech recognition, speech enhancement,
  and self-supervised learning representation,''
\newblock in {\em Interspeech}, 2022.

\bibitem{ccetin2006analysis}
{\"O}zg{\"u}r {\c{C}}etin and Elizabeth Shriberg,
\newblock ``Analysis of overlaps in meetings by dialog factors, hot spots,
  speakers, and collection site: Insights for automatic speech recognition,''
\newblock in {\em InterSpeech}, 2006.

\bibitem{delcroix2018single}
Marc Delcroix et~al.,
\newblock ``Single channel target speaker extraction and recognition with
  {Speaker Beam},''
\newblock in {\em IEEE ICASSP}, 2018.

\bibitem{vzmolikova2019speakerbeam}
Kate{\v{r}}ina {\v{Z}}mol{\'\i}kov{\'a} et~al.,
\newblock ``{SpeakerBeam}: Speaker aware neural network for target speaker
  extraction in speech mixtures,''
\newblock {\em IEEE Journal of Selected Topics in Signal Processing}, vol. 13,
  no. 4, 2019.

\bibitem{medennikov2020target}
Ivan Medennikov et~al.,
\newblock ``Target-speaker voice activity detection: a novel approach for
  multi-speaker diarization in a dinner party scenario,''
\newblock in {\em InterSpeech}, 2020.

\bibitem{wang2018voicefilter}
Quan Wang et~al.,
\newblock ``{VoiceFilter}: Targeted voice separation by speaker-conditioned
  spectrogram masking,''
\newblock in {\em InterSpeech}, 2019.

\bibitem{Chi2022XLMECL}
Zewen Chi et~al.,
\newblock ``{XLM-E}: Cross-lingual language model pre-training via electra,''
\newblock in {\em ACL}, 2022.

\bibitem{graves2006connectionist}
Alex Graves et~al.,
\newblock ``Connectionist temporal classification: labelling unsegmented
  sequence data with recurrent neural networks,''
\newblock in {\em Proceedings of the 23rd international conference on Machine
  learning}, 2006, pp. 369--376.

\bibitem{perez2018film}
Ethan Perez, Florian Strub, et~al.,
\newblock ``{FiLM}: Visual reasoning with a general conditioning layer,''
\newblock in {\em AAAI}, 2018.

\bibitem{ba2016layer}
Jimmy~Lei Ba, Jamie~Ryan Kiros, and Geoffrey~E Hinton,
\newblock ``Layer normalization,''
\newblock {\em ArXiv}, 2016.

\bibitem{pilault2020conditionally}
Jonathan Pilault, Christopher Pal, et~al.,
\newblock ``Conditionally adaptive multi-task learning: Improving transfer
  learning in {NLP} using fewer parameters \& less data,''
\newblock in {\em ICLR}, 2020.

\bibitem{cosentino2020librimix}
Joris Cosentino et~al.,
\newblock ``{LibriMix}: An open-source dataset for generalizable speech
  separation,''
\newblock {\em ArXiv}, 2020.

\bibitem{carletta2005ami}
Jean Carletta et~al.,
\newblock ``The {AMI} meeting corpus: A pre-announcement,''
\newblock in {\em International workshop on machine learning for multimodal
  interaction}, 2005.

\bibitem{librilight}
J.~{Kahn} et~al.,
\newblock ``{Libri-Light}: A benchmark for {ASR} with limited or no
  supervision,''
\newblock in {\em IEEE ICASSP}, 2020.

\bibitem{kanda2021large}
Naoyuki Kanda et~al.,
\newblock ``Large-scale pre-training of end-to-end multi-talker {ASR} for
  meeting transcription with single distant microphone,''
\newblock in {\em InterSpeech}, 2021.

\bibitem{watanabe2020chime}
Shinji Watanabe et~al.,
\newblock ``Chime-6 challenge: Tackling multispeaker speech recognition for
  unsegmented recordings,''
\newblock in {\em Proc. The 6th International Workshop on Speech Processing in
  Everyday Environments (CHiME 2020)}, 2020, pp. 1--7.

\bibitem{kahn2020libri}
Jacob Kahn et~al.,
\newblock ``Libri-light: A benchmark for asr with limited or no supervision,''
\newblock in {\em IEEE ICASSP}, 2020.

\bibitem{wang2021voxpopuli}
Changhan Wang et~al.,
\newblock ``{VoxPopuli}: A large-scale multilingual speech corpus for
  representation learning, semi-supervised learning and interpretation,''
\newblock in {\em ACL/IJCNLP}, 2021.

\bibitem{GigaSpeech2021}
Guoguo Chen et~al.,
\newblock ``Gigaspeech: An evolving, multi-domain asr corpus with 10,000 hours
  of transcribed audio,''
\newblock in {\em Interspeech}, 2021.

\bibitem{dehak2010front}
Najim Dehak et~al.,
\newblock ``Front-end factor analysis for speaker verification,''
\newblock {\em IEEE TASLP}, 2010.

\bibitem{snyder2018x}
David Snyder et~al.,
\newblock ``{X-vectors}: Robust {DNN} embeddings for speaker recognition,''
\newblock in {\em IEEE ICASSP}, 2018.

\bibitem{desplanques2020ecapa}
Brecht Desplanques et~al.,
\newblock ``{ECAPA-TDNN}: Emphasized channel attention, propagation and
  aggregation in {TDNN} based speaker verification,''
\newblock {\em InterSpeech}, 2020.

\bibitem{Peddinti2015ATD}
Vijayaditya Peddinti, Daniel Povey, and Sanjeev Khudanpur,
\newblock ``A time delay neural network architecture for efficient modeling of
  long temporal contexts,''
\newblock in {\em InterSpeech}, 2015.

\bibitem{Hu2020SqueezeandExcitationN}
Jie Hu, Li~Shen, Samuel Albanie, Gang Sun, and Enhua Wu,
\newblock ``Squeeze-and-excitation networks,''
\newblock {\em IEEE Transactions on Pattern Analysis and Machine Intelligence},
  vol. 42, 2020.

\bibitem{nagrani2017voxceleb}
Arsha Nagrani, Joon~Son Chung, and Andrew Zisserman,
\newblock ``{VoxCeleb}: A large-scale speaker identification dataset,''
\newblock in {\em InterSpeech}, 2017.

\bibitem{chung2018voxceleb2}
Joon~Son Chung, Arsha Nagrani, and Andrew Zisserman,
\newblock ``{VoxCeleb2}: Deep speaker recognition,''
\newblock in {\em InterSpeech}, 2018.

\bibitem{yu2017permutation}
Dong Yu, Morten Kolb{\ae}k, Zheng-Hua Tan, and Jesper Jensen,
\newblock ``Permutation invariant training of deep models for
  speaker-independent multi-talker speech separation,''
\newblock in {\em IEEE ICASSP}, 2017.

\bibitem{pearson1901liii}
Karl Pearson,
\newblock ``On lines and planes of closest fit to systems of points in space,''
\newblock {\em The London, Edinburgh, and Dublin philosophical magazine and
  journal of science}, 1901.

\end{thebibliography}

\end{document}